\documentclass[11pt,a4paper]{article}
\pdfoutput=1
\usepackage{graphicx}
\usepackage{jcappub}
\usepackage{indentfirst}

\usepackage{braket}
\usepackage{xcolor}
\usepackage{geometry}
\usepackage{verbatim}
\usepackage{subcaption}
\usepackage{csquotes}

\newcommand{\Tr}{\rm{Tr}}

\title{Bose-Einstein Condensates as Gravitational Wave Detectors}

\author[a,b]{Matthew P. G. Robbins,}
                                            \author[a,b]{Niayesh Afshordi}
                                            \author[a,b]{and Robert B. Mann}

                                            \affiliation[a]{Department of Physics and Astronomy, University of Waterloo,\\
                                            Waterloo, ON, Canada, N2L 3G1}
                                            \affiliation[b]{Perimeter Institute for Theoretical Physics,\\
                                            31 Caroline Street North, Waterloo, ON, Canada, N2L 2Y5}

\emailAdd{mrobbins@perimeterinstitute.ca}
\emailAdd{nafshordi@perimeterinstitute.ca}
\emailAdd{rbmann@uwaterloo.ca}

\abstract{We investigate a Bose-Einstein condensate (BEC) as a gravitational wave detector, and study its sensitivity by optimizing the properties of the condensate and the measurement duration. 
We show that detecting kilohertz gravitational waves is limited by current experimental techniques in squeezing BEC phonons. Future improvements in technology to squeeze BEC states can make them competitive detectors for gravitational waves of astrophysical and/or cosmological origin. 
}

\keywords{gravitational wave detectors, gravitational waves / experiments, quantum field theory on curved space}

\arxivnumber{arXiv:1811.04468 [quant-ph]}

\begin{document}

\maketitle
\flushbottom
\section{Introduction}

With the direct detection of gravitational waves, an entirely new avenue of studying the universe has opened. 
It is incumbent upon the scientific community to consider both the theoretical aspects of this discovery and to propose new means of detecting gravitational waves.

At the present time, gravitational wave detection is primarily done with interferometers. A drawback is that they are sensitive to only a small range of frequencies, with LIGO being most sensitive around 100-300 Hz,   allowing it to detect stellar-mass inspiralling black holes and neutron stars \cite{LIGO2016}. Gravitational wave detectors with sensitivities at lower frequencies have been proposed, which will allow the large-mass black hole binary  inspirals to be studied \cite{Moore2015}. At kilohertz frequencies, the creation of magnetars \cite{Stella2005} and signals from more neutron star mergers \cite{Andersson2003,Andersson2011} can also be observed with improved sensitivity.  Detection of kHz gravitational waves will have wide-ranging implications in further understanding the universe, such as constraining the equation of state of neutron stars \cite{Andersson1998}.

Interferometers involving cold atoms have previously been suggested as a means to detect low frequency gravitational waves \cite{Dimopoulos2009,Hogan2010}. Recently, a novel suggestion  \cite{Sabin2014}  making use of a Bose-Einstein condensate (BEC) as a high frequency gravitational wave detector has been proposed.   A zero temperature quasi (1+1)-dimensional BEC with fluctuating boundaries was considered in the presence of a gravitational wave with plus polarization (in the BEC frame), $h_+=\epsilon\sin\Omega t$, where $\epsilon$ is the amplitude of the gravitational wave and $\Omega$ is its frequency. It was shown that gravitational waves are able to create phonons in the BEC, such that the number of particles produced was $\mathcal{O}(\epsilon^2)$. As it would be impractical to detect the phonon production directly, techniques of quantum metrology were suggested. By calculating the fidelity between phonon states, it was possible to determine the quantum Fisher information $H_{\epsilon}$ of the phonon state of the BEC, which characterizes the amount of information contained in the amplitude of the gravitational wave \cite{Braustein1994,Sabin2014}:
\begin{align}
M\braket{(\Delta\epsilon)^2}\geq\frac{1}{H_{\epsilon}}\ ,
\end{align}
where $M$ is the number of independent measurements of the system and $\braket{(\Delta\epsilon)^2}$ is the mean-square-error in the amplitude of the gravitational wave. Assuming that the phonons are in squeezed two-mode states, it was demonstrated that, with a suitable number of measurements of the fidelity between phonons interacting with a gravitational wave of amplitude $\epsilon$ and a gravitational wave of amplitude $\epsilon+d\epsilon$, the strain sensitivity $\sqrt{\langle \Delta \epsilon^2 \rangle}$ was able to exceed that of LIGO at frequencies in the kilohertz regime and above.

We offer here a more complete perspective by considering zero-temperature (3+1)-dimensional BECs, where
$h_+=\epsilon e^{-t^2/\tau^2}\sin(\Omega t)$ is used to model
 the incoming gravitational wave; the quantity $\tau$ captures the duration of a single measurement of the BEC. We treat the phonons as being in single-mode squeezed states and examine whether current techniques of squeezing phonons and producing BECs are sufficient in order to use the condensate as a gravitational wave detector.

In Section \ref{sec: BECs curved}, we introduce BECs in a curved spacetime and derive the Euler-Lagrange equation for the phonons being influenced by the gravitational waves. Then, in Section \ref{sec: GW detection}, we apply techniques in quantum metrology to estimate the sensitivity for the detection of gravitational waves. In Section \ref{sec: Experiment}, we consider methods of constructing a BEC gravitational wave detector.  We show that if the phonons are restricted to obey a linear dispersion relation, then the amount of phonon squeezing is the dominant limiting factor. We also address the damping present in the condensate at $T=0$ and comment on how to increase the sensitivity of the condensate to gravitational waves. Section \ref{sec: Conclusion} presents our conclusions.

\section{Bose-Einstein condensates in a curved background}
\label{sec: BECs curved}

We will now derive the equation of motion for the phonons as well as its Bogoliubov coefficients. Related derivations can be found in \cite{Nicolis2014,Sabin2014,Fagnocchi2010}.

The Lagrangian for a BEC in a curved background is
\begin{align}
\mathcal{L}=g^{\mu\nu}\partial_{\mu}\phi\partial_{\nu}\phi^{\dagger}-m^2|\phi|^2-U(|\phi|^2)
\label{eq: L}
\end{align}
where $m$ is the mass of the atoms of the BEC, $\phi$ is the field,  $U(|\phi|^2)=\lambda|\phi|^4 > 0$ describes the interaction of the BEC.\footnote{ As this describes the generalization of a BEC in curved spacetime, we note that it will only be accurate for $na^3 =\frac{(\lambda c_s)^2}{(4\pi)^3 c^2} \ll1$ as this is the regime of validity for the Gross-Pitaevskii equation, where $n$ is the number density, $a$ is the s-wave scattering wavelength, and $c_s$ is the speed of sound defined below in Equation (\ref{eq: cs}) .} Let us write $\phi=\hat{\phi}e^{i\chi}$ (with real $\hat{\phi}$ and $\chi$) and assume that the BEC is homogeneous. 
We want to determine the $\hat{\phi}$ that extremizes $\mathcal{L}$. Differentiating with respect to $\hat{\phi}$, we find that the extremum occurs at
\begin{align}
\hat{\phi}^2=\frac{1}{2\lambda}\left[\partial^{\mu}\chi\partial_{\mu}\chi-m^2\right]\ .
\label{eq: phi2}
\end{align}
Inserting (\ref{eq: phi2}) into (\ref{eq: L}) and writing $\chi=\kappa t+\pi$ (where $\pi\in\Re$ is the pseudo-Goldstone boson, describing the BEC acoustic  perturbations or phonons), the action becomes
\begin{equation}
\begin{aligned}
S&=\int\frac{d^4x}{4\lambda}\sqrt{-g}\left\{(\kappa \delta_0^{\nu}+\partial^{\nu}\pi)(\kappa \delta_0^{\mu}+\partial^{\mu}\pi)g_{\mu\nu}-m^2\right\}^2\ ,\\
&=\int\frac{d^4x}{4\lambda}\sqrt{-g}\left[\kappa ^2g_{00}+2\kappa\dot{\pi}g_{00}+\kappa \partial^i\pi g_{0i}+|\dot{\pi}|^2g_{00}+\partial^i\pi\partial^j\pi g_{ij}+2\partial^i\pi\dot{\pi}g_{i0}-m^2\right]^2\ .
\end{aligned}
\end{equation}
Let us work in the TT-gauge and take $g_{\mu\nu}=\eta_{\mu\nu}+h_{\mu\nu}$, where
\begin{align}
h_{\mu\nu}=\begin{pmatrix}
0 & 0 & 0 & 0\\
0 & h_+(t) & h_\times(t) & 0\\
0 & h_\times(t) & -h_+(t) & 0\\
0 & 0 & 0 & 0
\end{pmatrix}\ ,
\end{align}
describes a gravitational wave propagating in the $z$-direction and $h_+$ and $h_\times$ are the two polarizations of the gravitational wave.\footnote{ We assume that the trapped particles, representing the boundary of the BEC box, move on geodesics. Therefore, starting at rest, they will not see the gravitational waves to linear order.} Using $\sqrt{-g}\sim 1+\mathcal{O}(h_{\mu\nu}h^{\mu\nu})$ and a $(+,-,-,-)$ signature, we can expand in terms of $\pi$ to find 
\begin{align}
S&=\int\frac{d^4x}{4\lambda}\left[\kappa^2+2\kappa\dot{\pi}+|\dot{\pi}|^2+\partial^i\pi\partial^j\pi g_{ij}-m^2\right]^2\ ,\label{eq: S1}\\
&\approx\int\frac{d^4x}{4\lambda}\left[|\dot{\pi}|^2(6\kappa^2-2m^2)+(2\kappa^2-2m^2)(\eta_{ij}+h_{ij})\partial^i\pi\partial^j\pi\right]\ ,
\label{eq: S2}
\end{align}
where  the first-order terms can be written as a total derivative that integrates to zero on the boundary
and we assume that the higher-order terms can be neglected.
Therefore, the Euler-Lagrange equation is
\begin{align}
\ddot{\pi}+c_s^2(\eta_{ij}+h_{ij})\partial^i\partial^j\pi+c_s^2(\partial^jh_{ij})\partial^i\pi=0\ ,
\end{align}
where
\begin{align}
 c^2_s \equiv \frac{\kappa^2-m^2}{3\kappa^2-m^2}\ .
 \label{eq: cs}
\end{align}
In Appendix \ref{sec: appendix}, instead of making the low-energy (or adiabatic) approximation (\ref{eq: phi2}), we solve for the full linear perturbations (for $\hat{\phi}$ and $\pi$). This shows the above derivation is valid as long as $\omega \ll \mu$, where $\mu = \kappa -m \approx m c^2_s$ is the chemical potential of the BEC.

For simplicity, we will assume that $h_{\times}=0$. We model the plus polarization as $h_+=\epsilon e^{-t^2/\tau^2}\sin\Omega t$ (ignoring its spatial dependence\footnote{This is justified when the speed of sound is much smaller than that of gravitational waves, $c_s \ll 1$.}), where $\epsilon$ is the amplitude of the gravitational wave, $\tau$ captures the duration of a single measurement of the gravitational wave, and $\Omega$ is the frequency of the wave. 
 Noting that  $(\partial_ih^{ij})=0$ in the TT gauge and using the ansatz $\pi\propto e^{ik\cdot x}\psi(t)$, we find
\begin{equation}
\begin{aligned}
(3\kappa^2-m^2)\ddot{\psi}+(\kappa^2-m^2)(\eta_{ij}+h_{ij})k^ik^j\psi=0 
\end{aligned}
\end{equation}
up to a normalization of $\psi$.
Thus, with $k^1=k_x$, $k^2=k_y$, and $k^3=k_z$,
\begin{align}
(3\kappa^2-m^2)\ddot{\psi}+(\kappa^2-m^2)|k|^2\big[1+\tilde{\epsilon} e^{-t^2/\tau^2}\sin(\Omega t) \big]\psi=0\ .
\label{eq: equation}
\end{align}
where $\tilde{\epsilon}=\frac{(k_x^2-k_y^2)}{|k|^2}\epsilon$. We can then rewrite Equation (\ref{eq: equation}) as
\begin{align}
\ddot{\psi}+[1+\tilde{\epsilon} e^{-t^2/\tau^2}\sin(\Omega t)]c_s^2k^2\psi=0\ ,
\label{eq: SHO}
\end{align}

Let us now look at the validity of Equation (\ref{eq: SHO}), which was determined by neglecting higher-order terms in Equation (\ref{eq: S1}). The third-order terms in the Lagrangian are
\begin{align}
\mathcal{L}_3=\frac{\kappa}{\lambda}\left[|\dot{\pi}|^3+\dot{\pi}g_{ij}\partial^i\pi\partial^j\pi\right]\ .
\label{eq: L3}
\end{align}
For $\kappa\approx m$, we have $c_s^2\approx\frac{\kappa^2-m^2}{2m^2}\ll1$, so $4m^2|\dot{\pi}|^2\approx 2m^2c_s^2|\nabla\pi|^2$ on average. Now, comparing Equations (\ref{eq: S2}) and (\ref{eq: L3}),  we see that linear linear theory is only valid when
\begin{equation}
c^{-1}_s|\dot{\pi}| \sim |\nabla\pi| \ll m c_s\ . \label{linear}
\end{equation}

Solving (\ref{eq: SHO})  perturbatively by writing $\psi=\psi^{(0)}+\tilde{\epsilon}\psi^{(1)}$  yields
\begin{align}
\ddot{\psi}^{(0)}+\omega^2\psi^{(0)}&=0\ ,\\
\ddot{\psi}^{(1)}+\omega^2\psi^{(1)}&=-\omega^2e^{-t^2/\tau^2}\sin(\Omega t)\psi^{(0)}\ ,
\end{align}
which has the solutions
\begin{align}
\psi^{(0)}(t)&=C_1^{(0)}e^{i\omega t}+C_2^{(0)}e^{-i\omega t}\ ,\label{eq: psi0}\\
\psi^{(1)}(t)&=C_1^{(1)}e^{i\omega t}+C_2^{(1)}e^{-i\omega t} -\int_{-\infty}^{\infty}dt_1
\omega^2 e^{-t_1^2/\tau^2}\sin(\Omega t_1) G(t,t_1)\psi^{(0)}(t_1)
\ ,\label{eq: psi1}
\end{align}
where
\begin{align}
G(t,t_1)=\frac{\sin[\omega(t-t_1)]}{\omega}\Theta(t-t_1)\ ,
\end{align}
is the Green's function of a harmonic oscillator, while $\Theta$ represents the Heaviside function.  Combining our ansatz of $\pi\propto e^{ik\cdot x}\psi(t)$ with Equations (\ref{eq: psi0}) and (\ref{eq: psi1}), we find  
\begin{align}
\pi(x,t)\propto e^{ikx}\left[e^{-i\omega t}+\frac{\sqrt{\pi}\tilde{\epsilon}\omega\tau}{4}e^{-\frac{1}{4}\left(\Omega+2\omega\right)^2\tau^2}\left(e^{2\omega\Omega\tau^2}-1\right)e^{i\omega t}\right]\ .
\label{eq: pi}
\end{align}
As we are working in curved spacetime with a single mode, we can write $\pi(x,t)\propto e^{ikx}\left[\alpha e^{-i\omega t}+\beta e^{i\omega t}\right]$, where $\alpha$ and $\beta$ are Bogoliubov coefficients. This immediately yields
\begin{align}
\alpha&=1\ ,\label{eq: alpha}\\
\beta&=\frac{\tilde{\epsilon}\sqrt{\pi}\omega}{4}\tau e^{-(\Omega+2\omega)^2\tau^2/4}\left(e^{2\omega\Omega\tau^2}-1\right) \ , \label{eq: beta}
\end{align}
from Equation (\ref{eq: pi}). 

Let us briefly comment on the large and small $\tau$ limits. We note that $\beta\to0$ as $\tau\to0$. Intuitively, this makes sense because a vanishing measurement duration implies that no information about the gravitational wave would be obtained. We further discuss the information that can be acquired in Section \ref{sec: GW detection}. For $\tau\to\infty$ we recover (as expected \cite{Sabin2014}) $\beta\to0$.  We note that for
non-geodesic boundaries \cite{Sabin2014},  non-zero Bogoliubov coefficients $\alpha_{nm}$ and $\beta_{nm}$ result for modes $n\neq m$. However in our case coefficients with $n\neq m$ are zero since we assume non-interacting modes. 

As we are considering an odd function for the gravitational waveform, we find $\alpha=1$.
 If, for example, we had instead considered an even function $h_+=\epsilon e^{-t^2/\tau^2}\cos(\Omega t)$, then $\alpha$ would include a non-zero imaginary $\mathcal{O}(\epsilon)$ term. Such effects are necessary to consider if a BEC gravitational wave detector were constructed; we shall neglect this additional effect  henceforth in order to keep the discussion as simple as possible.

As $\beta\propto\epsilon$, the number of particles created is $\mathcal{O}(\epsilon^2)\ll1$, so it is not straightforward to directly count the number of particles created. We will therefore adapt the quantum metrology procedure of \cite{Sabin2014b} to determine the amplitude of the gravitational wave.

\section{Detection of gravitational waves via quantum metrology}
\label{sec: GW detection}

Quantum metrology is the study of making precision measurements by exploiting quantum mechanical properties, rather than solely relying on classical measurements of a system. This can be used, for example,  to overcome  shot noise in a detector \cite{Giovannetti2006,Nature2012,Giovannetti2012}. Another advantage is that quantum metrology can be used to estimate a parameter $\theta$ that is not an operator observable of a system. This is done by determining how an infinitesimal change of the parameter affects the statistical distance between two quantum states, thereby defining their distinguishability (fidelity) and quantum Fisher information.

An estimate in the error in the measurement of $\theta$ is obtained from the quantum Fisher information \cite{Braustein1994}
\begin{align}
H(\theta)=\frac{8\bigg(1-\sqrt{F(\rho_{\theta},\rho_{\theta+d\theta})}\bigg)}{d\theta^2}
\label{eq: QFI}
\end{align}
where $F(\rho_{\theta},\rho_{\theta+d\theta})$ is the fidelity between the states $\rho_{\theta}$ and $\rho_{\theta+d\theta}$. The fidelity between two states $\rho',\rho''$ is defined as \cite{Uhlmann1976,Jozsa1994}
\begin{align}
F(\rho',\rho'')=\left[\Tr\sqrt{\rho'\sqrt{\rho''}\rho'}\right]^2\ ,
\end{align}
which describes the overlap between $\rho'$ and $\rho''$. When both $\rho'$ and $\rho''$ correspond to Gaussian states, it is often easier to use covariance matrices. In this case, the covariance matrix for a Gaussian state is $\sigma_{mn}=\frac{1}{2}\braket{X_mX_n+X_nX_m}-\braket{X_m}\braket{X_n}$, where $X_{2n-1}=\frac{1}{\sqrt{2}}(a_n+a_n^{\dagger})$, $X_{2n}=\frac{1}{\sqrt{2}i}(a_n-a_n^{\dagger})$, and $a_n,a_n^{\dagger}$ are the creation and annihilation operators. Note that this normalization convention is different from what was used in \cite{Sabin2014,Sabin2014b}. Suppose that $M$ independent measurements are done to determine $\theta$. Then \begin{align}
\braket{(\Delta\theta)^2}\geq\frac{1}{MH(\theta)}
\label{eq: max error}
\end{align}
is the minimum error in measuring $\theta$  \cite{Braustein1994,Sabin2014}.

We will now use the quantum Fisher information to estimate the minimum error in the amplitude of the gravitational wave. We will restrict ourselves to the case in which $\braket{X_i}=0$. First, consider two Gaussian states described by the covariance matrices $\sigma_A,\sigma_B$. Let us define
\begin{subequations}
\begin{align}
\Delta&=\det[\mathcal{\sigma}_A+\mathcal{\sigma}_B]\ ,\label{eq: Delta}\\
\Lambda&=2^{2n}\det\left[\mathcal{\sigma}_A+\frac{i}{2}J\right]\det\left[\mathcal{\sigma}_B+\frac{i}{2}J\right]\ ,\label{eq: Lambda}\\
J&=\bigoplus_{k=1}^n\begin{pmatrix}
0 & 1\\
-1 & 0
\end{pmatrix}\ ,
\end{align}
\label{eq: DLJ}
\end{subequations}
where $n$ is the number of modes. Now, the fidelity between the two covariance matrices of two single-mode states is given by \cite{Marian2012}
\begin{align}
F(\sigma_A,\sigma_B)=\frac{1}{\sqrt{\Delta+\Lambda}-\sqrt{\Lambda}}\ .
\label{eq: F}
\end{align}

Consider preparing the phonons in the BEC in a squeezed Gaussian single-mode state, described by an initial covariance matrix $\sigma(0)$ (at zero-temperature) \cite{Ferraro2005}:
\begin{align}
\sigma(0)=\frac{1}{2}\begin{pmatrix}
\cosh(2r)+\cos(\phi)\sinh(2r) & -\sin(\phi)\sinh(2r) \\
-\sin(\phi)\sinh(2r) &\cosh(2r)-\cos(\phi)\sinh(2r)
\end{pmatrix}\ ,
\label{eq: sigma 0}
\end{align}
where $r$ is the squeezing parameter and $\phi$ is the squeezing angle. When a gravitational wave passes by the BEC, it affects the phonons by transforming its covariance matrix to $\sigma_\ell(\tilde{\epsilon})$ where \cite{Fuentes2014}
\begin{align}
\sigma_\ell(\tilde{\epsilon})=\mathcal{M}_{\ell\ell}(\tilde{\epsilon})\sigma(0)\mathcal{M}_{\ell\ell}(\tilde{\epsilon})+\sum_{j\neq \ell}\mathcal{M}_{\ell j}(\tilde{\epsilon})\mathcal{M}_{\ell j}^T(\tilde{\epsilon})\ ,
\label{eq: sigma k}
\end{align}
with $\ell$   the mode number of the phonon and
\begin{align}
\mathcal{M}_{mn}(\tilde{\epsilon})=\begin{pmatrix}
\Re[\alpha_{mn}-\beta_{mn}] & \Im[\alpha_{mn}-\beta_{mn}] \\
-\Im[\alpha_{mn}-\beta_{mn}] & \Re[\alpha_{mn}+\beta_{mn}] 
\end{pmatrix} \, .
\label{eq: M kn}
\end{align}
Our Bogloliubov coefficients in Equations (\ref{eq: alpha}) and (\ref{eq: beta}) do not couple different modes, so we note that  $\alpha_{mn}=\alpha\delta_{mn}$ and $\beta_{mn}=\beta\delta_{mn}$.

As shown in \cite{Sabin2014b}, Equation (\ref{eq: QFI}) can be written as
\begin{align}
H(\epsilon)=2\left(\sigma_{11}^{(0)}\sigma_{22}^{(2)}+\sigma_{11}^{(2)}\sigma_{22}^{(0)}-2\sigma_{12}^{(1)}\sigma_{22}^{(1)}\right)+\frac{1}{2}\left(\sigma_{11}^{(1)}\sigma_{22}^{(1)}-2\sigma_{12}^{(1)}\sigma_{12}^{(1)}\right),
\label{eq: QFI2}
\end{align}
where $\sigma_{ij}^{(n)}$ is the $ij$ matrix element in an expansion of the covariance matrix
\begin{align}
\sigma_{ij}(\epsilon)=\sigma_{ij}^{(0)}+\sigma_{ij}^{(1)}\epsilon+\sigma_{ij}^{(2)}\epsilon^2+\mathcal{O}(\epsilon^3) 
\label{eq: sigma ij}
\end{align}
in powers of $\epsilon$, 
and we have assumed that the amount of displacement of the squeezed state is zero. Therefore, using Equations (\ref{eq: sigma 0})-(\ref{eq: sigma ij}), we find

\begin{equation}
\begin{aligned}
\frac{1}{M\braket{(\Delta\tilde{\epsilon})^2}}\leq\frac{\pi\omega^2\tau^2}{64}R\left(e^{-\frac{(\Omega -2
   \omega )^2\tau^2}{4 }}-e^{-\frac{(\Omega+2 \omega )^2\tau^2}{4}}\right)^2
      \label{eq: delta epsilon 1 mode}
\end{aligned}\ ,
\end{equation}
where
\begin{align}
R \equiv \sinh ^2(2 r) (6\sin^2\phi-2)+\cosh (4 r)+1\ .
   \label{eq: M'}
\end{align}

Equation (\ref{eq: delta epsilon 1 mode}) describes a BEC possessing a single mode, but we can further exploit all modes of the BEC to improve the sensitivity. From $\tilde{\epsilon}=\frac{k_x^2-k_y^2}{k^2}\epsilon$, we have $\braket{(\Delta\tilde{\epsilon}_{\vec{k}})^2}=\left(\frac{k_x^2-k_y^2}{k^2}\right)^2\braket{(\Delta\epsilon_{\vec{k}})^2}$, where $\braket{(\Delta\epsilon_{\vec{k}})^2}$ is the error in $\epsilon$ for mode $\vec{k}$. Averaging over the solid angle, $\frac{1}{4\pi}\int d\theta d\varphi \braket{(\Delta\tilde{\epsilon}_{\vec{k}})^2}\sin\theta=\frac{4}{15}\braket{(\Delta\epsilon_{\vec{k}})^2}$. Substituting this average into $\braket{(\Delta\tilde{\epsilon}_{\vec{k}})^2}$, we find
\begin{align}
\frac{1}{\braket{(\Delta \epsilon)^2}}_{tot}=\sum_{\vec{k}} \left\langle \left(\frac{k_x^2-k_y^2}{k^2}\right)^2 \right\rangle \frac{1}{\braket{(\Delta\tilde{\epsilon}_{\vec{k}})^2}}=\frac{4}{15}\sum_{\vec{k}}\frac{1}{\braket{(\Delta \tilde{\epsilon}_{\vec{k}})^2}}\ ,
\label{eq: Delta epsilon discrete}
\end{align}
where $\braket{(\Delta \epsilon)^2}_{tot}$ is the total error in the measurement of the amplitude of the gravitational wave. For a large number of single-mode states and assuming that the modes are non-interacting, we can convert Equation (\ref{eq: Delta epsilon discrete}) into an integral. With $\omega=c_sk$, assuming the BEC has a volume of $L^3$, $k=\sqrt{\left(\frac{n_x\pi}{L}\right)^2+\left(\frac{n_y\pi}{L}\right)^2+\left(\frac{n_z\pi}{L}\right)^2}$, and using spherical coordinates, we have
\begin{equation}
\begin{aligned}
\frac{1}{\braket{(\Delta \epsilon)^2}}_{tot}&\lesssim\frac{\pi ^4M c_s^2R\tau^2}{480 L^2}\int_0^{\infty} n^4 e^{-\frac{(2 \pi c_s n+L \Omega )^2\tau^2}{2 L^2}} \left(e^{\frac{2 \pi 
   c_s n \Omega\tau^2 }{  L}}-1\right)^2 dn\ ,\\
   &=\frac{ML^3 e^{-\frac{1}{2} \tau ^2 \Omega ^2}R \left(e^{\frac{\tau ^2 \Omega ^2}{2}}
   \left(\tau ^4 \Omega ^4+6 \tau ^2 \Omega ^2+3\right)-3\right)}{7680 \sqrt{2 \pi } c_s^3 \tau ^3}\ ,      \label{eq: Delta epsilon continuous}
\end{aligned}
\end{equation}
neglecting $\mathcal{O}(\epsilon^2)$ terms and  only considering the cases in which the integrand is peaked at $n\gg 1$ (corresponding to gravitational wave frequency much bigger than the lowest acoustic harmonic of the BEC, $\Omega \gg \frac{2\pi c_s}{L}$).

Now, for a total observation time of $t_{obs}$, we can approximately run $M \sim t_{obs}/\tau$ separate measurements of the BEC state. Therefore,
\begin{equation}
\begin{aligned}
\braket{(\Delta \epsilon)^2}_{tot}&\geq\frac{7680\sqrt{2 \pi } c_s^3  {\tau ^4} e^{\frac{\tau ^2 \Omega ^2}{2}}}{L^3
   t_{obs} R\left(e^{\frac{\tau ^2 \Omega ^2}{2}} \left(\tau ^4 \Omega ^4+6 \tau ^2
   \Omega ^2+3\right)-3\right)}\ .
   \end{aligned}
   \label{eq: M}
\end{equation}
Note that $R$ is maximized at $\phi=\pi/2$, such that $R_{\rm max}=3 \cosh (4 r)-1$. We point out that squeezing at specific angles is physical and has previously been done \cite{Chelkowski2005,Johnsson2013}. 

For $\Omega\tau\ll1$,
\begin{align}
\braket{(\Delta \epsilon)^2}_{tot}\geq\frac{1024 \sqrt{2\pi} c_s^3 \tau ^2}{L^3 t_{obs}\Omega ^2R}+\mathcal{O}(\Omega^4\tau^4)\ ,
\end{align}
indicating that shorter (individual) measurement times, $\tau$, for a fixed total observation time $t_{obs}$, will maximize the sensitivity of the BEC to an incoming gravitational wave. However, $\tau$ cannot be made arbitrarily short;  we shall briefly discuss this in Section \ref{sec: Experiment}. 

 Let us now investigate the maximum squeezing of the system. From Equation (\ref{eq: S2}), for $4|\dot{\pi}|^2\approx 3c_s^2|\nabla\pi|^2$, which is true on average, the Hamiltonian is
\begin{align}
H=\frac{7m^2}{4\lambda}\int d^3x|\nabla\pi|^2\sim\frac{7m^2}{4\lambda}L^3 |\nabla\pi|^2 
\label{eq: H1}
\end{align}
where $V$ is the volume of the condensate.
In the ground state of the system, we would be able to write the Hamiltonian as
\begin{align}
H=L^3\int\frac{d^3k}{(2\pi)^2}\frac{1}{2}\omega\sim\frac{L^3}{32\pi^2}k^4c_s\ ,
\label{eq: H2}
\end{align}
where $k$ is the maximum effective wavenumber obeying the linearity condition (\ref{linear}): $\braket{|\nabla\pi|^2}\ll m^2c_s^2$. Comparing Equations (\ref{eq: H1}) and (\ref{eq: H2}), we find $|\nabla\pi|^2\sim\frac{\lambda k^4 c_s}{56m^2\pi^2}$. If we squeeze the ground state, then we require $\frac{\lambda k^4 c_s}{ {56}m^2\pi^2}e^{2r}\ll m^2c_s^2$. Therefore, after re-inserting $\hbar$ and $c$, we find
\begin{align}
e^{2r}\ll \frac{56\pi^2}{\lambda}\left(\frac{m c}{\hbar k}\right)^4\left(\frac{c_s}{c}\right)\ .
\label{eq: e2r}
\end{align}
To calculate $\lambda$, we need only to consider the background, such that $\pi=0$. With
\begin{align}
T_{\mu\nu}=\frac{2}{\sqrt{-g}}\frac{\delta S}{\delta g^{\mu\nu}}\ ,
\end{align}
and
\begin{align}
S=\int d^4x\sqrt{-g}\mathcal{L}\ , 
\end{align}
where $\displaystyle{\mathcal{L}=\frac{1}{4\lambda}\left[g^{\mu\nu}\partial_{\mu}\chi\partial_{\nu}\chi-m^2\right]^2}$, we find
\begin{align}
T_{\mu\nu}=g_{\mu\nu}\mathcal{L}+4\left[g^{\mu'\nu'}\partial_{\mu'}\chi\partial_{\nu'}\chi-m^2\right]\partial_{\mu}\chi\partial_{\nu}\chi\ .
\end{align}
Let $\rho=T_{00}$ be the energy density. Then, with $\dot{\pi}=0$, $\nabla\pi\sim\dot{\pi}$, and $2m^2c_s^2\approx\kappa^2-m^2$, we see (again re-inserting $\hbar$ and $c$)
\begin{align}
\rho\approx\frac{1}{\lambda}\frac{m^4c_s^2c^3}{\hbar^3}\ .
\end{align}
Therefore, Equation (\ref{eq: e2r}) becomes
\begin{align}
e^{2r}\ll  \frac{56 \pi ^2 c_s^3 \rho }{\omega ^4 \hbar }
\label{eq: approx}
\end{align}
where we have also used $\omega=c_sk$. For a BEC  with number density $7\times10^{20}$ m${}^{-3}$ containing atoms of mass $10^{-25}$ kg and in the case of phonons of frequency $\frac{\omega}{2\pi} = 10^4$ Hz and speed $c_s=1.2$ cm/s, we have $r\lesssim27$.

\section{Experimental realization}
\label{sec: Experiment}

\subsection{Experimental Techniques}

Let us now analyze the experimental feasibility of using a BEC to detect gravitational waves. We first  note that position squeezing \cite{Lvovsky2014}
\begin{align}
s=-10\log_{10}\left(e^{-2r}\right)
\end{align}
is reported in decibels.
 Numerical simulations involving optomechanics and trichromatic lasers have been able to squeeze phonons by at least 7.2 dB \cite{Gu2014}, corresponding to a squeezing parameter of $r=0.83$. Phonons have been squeezed using second-order Raman scattering \cite{Hu1997,Garrett1997}, though this was in the presence of a crystal lattice. Even though we are interested in a BEC in a curved spacetime, it may still be possible to exploit this feature.

As shown in \cite{Jaksch1998}, the presence of an optical lattice potential implies that the flat space BEC Hamiltonian can be written as a Bose-Hubbard Hamiltonian.  In this case, a modification of the methods of squeezing phonons in crystal lattices, such as second order Raman scattering \cite{Hu1997} or pump-probe detection scheme \cite{Benatti2017} might potentially be used.  The current limitation on squeezing phonons arguably represents one of the greatest challenges for using a BEC as a gravitational wave detector. With $r=0.83$ and $\phi=\pi/2$, we see that $R\approx 41$. If phonons were squeezed 20 dB, corresponding to $r=2.3$,
 then $R\approx 1.5\times10^4$. An increase in the amount of squeezing would exponentially increase the sensitivity to the gravitational wave \cite{Hosten2016}. For simplicity, we have assumed that the BEC is cubic in shape with no external potential, but realistically, we note that only certain trap geometries and trapping potentials lend themselves to large (number) squeezing \cite{Johnsson2013}. 

Let us now consider the ratio $\frac{c_s^3}{L^3t_{obs}}$. Experiments have been done to create condensates with lengths on the order of tens of microns to mm \cite{Vengalattore2018,Barr2015,Greytak2000}. However, these lengths are only in a single direction, with the other length (in the case of quasi-two-dimensional BECs) much smaller. As shown in \cite{Andrews1997,Andrews1997Erratum}, speeds of sound in BECs were analyzed as a function of the density, with a speed of approximately 1.2 cm/s being obtained at a number density of $7\times10^{20}$ m${}^{-3}$. 

 For a gravitational wave of period $T$, sensitivity is optimized for $T\lesssim\tau\lesssim t_d$, where $t_d$ is the decoherence time of the phonons, which we discuss in more detail in Section \ref{sec: decoherence}. For gravitational waves in the kHz frequency range, the minimum time required is $\tau\gtrsim 10^{-3}$ s. One proposal in \cite{Sabin2014} is to use quantum dots to make measurements on the BEC in which they suggested using 1500 dots to make $10^6$ measurements per second. 
 
 \begin{figure}[t!]
\centering
        \includegraphics[width=\textwidth]{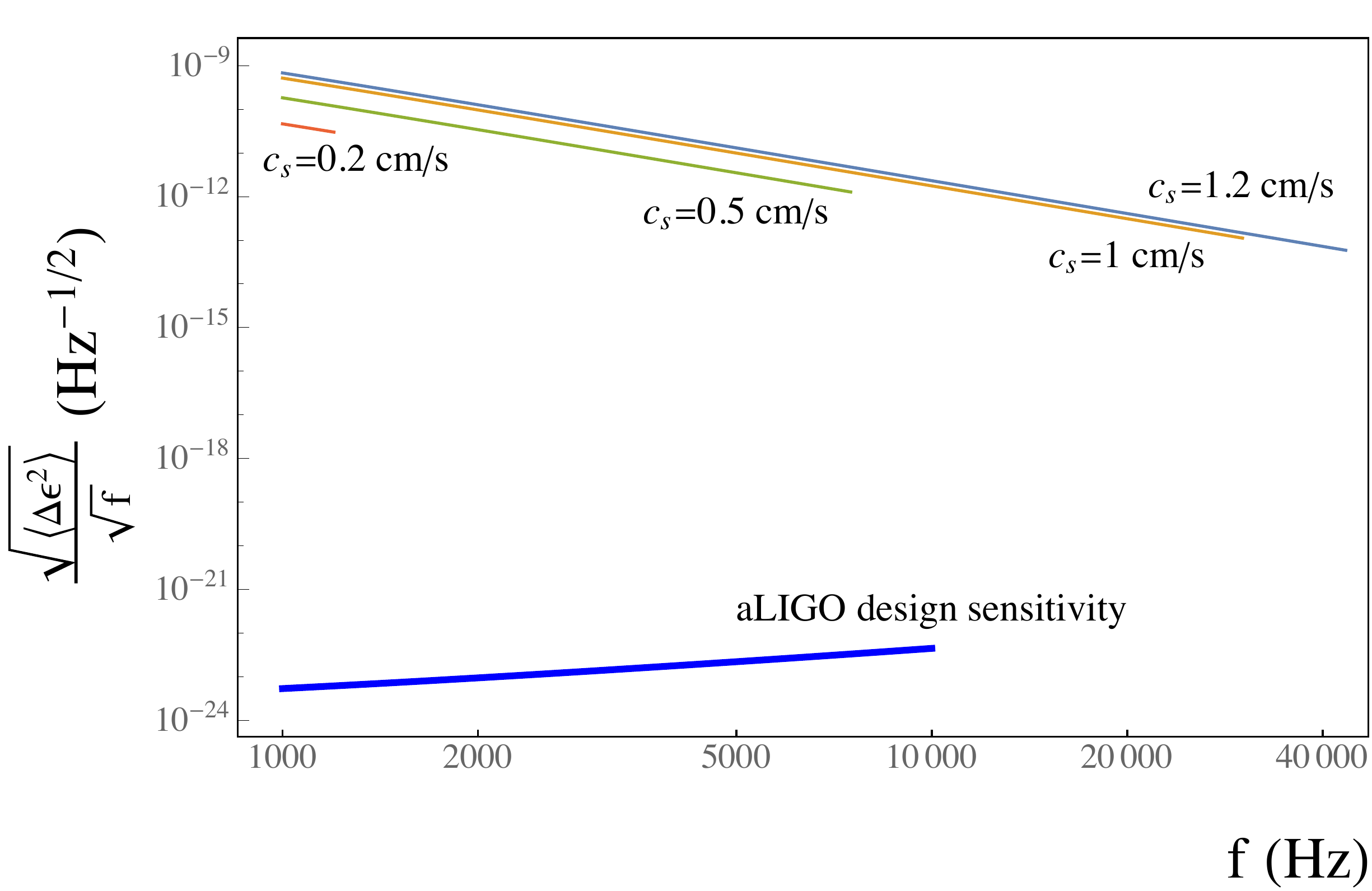}\hspace{0.5cm}
    \caption{Sensitivities of a BEC gravitational wave detector (top curve) using current experimental capabilities, where $f$ is the frequency of the gravitational wave. With $L=10^{-3}$ m, $t_{obs}=10^6$ s, $\tau=10^{-3}$ s, $\phi=\pi/2$, and $r=0.83$, a BEC is unable to observe kHz gravitational waves. The thick blue line corresponds to the maximum design sensitivity of aLIGO (extrapolated to $10^4$ Hz).  Note that phonons of frequency $\frac{f}{2}$ are necessary to detect gravitational waves of frequency $f$.}
    \label{fig: Current 1}
\end{figure}
 
 In figure \ref{fig: Current 1}, we assume that a BEC can be constructed with the best experimental/numerical parameters that have been achieved: modes are squeezed by 7.2 dB ($r=0.83$) \cite{Gu2014}, the BEC has a length of $L=10^{-3}$ m \cite{Vengalattore2018,Barr2015,Greytak2000},  the speed of sound of $c_s=1.2\times10^{-2}$ m/s  \cite{Andrews1997,Andrews1997Erratum}, the quadrature angle is $\phi=\pi/2$, and there is a total observational time of $10^6$ s. It is seen that our maximum sensitivity for a gravitational wave in the kHz regime is approximately $6\times10^{-14}$ Hz${}^{-1/2}$.  From Equation (\ref{eq: M}), though smaller speeds of sound will increase the sensitivity,  the available frequency range will also decrease because the chemical potential $\mu=m c_s^2$ becomes smaller. To observe a gravitational wave of $1$ kHz using atoms of $m=10^{-25}$ kg, the maximum sensitivity of $4\times10^{-11}$ Hz${}^{-1/2}$ occurs when $\mu=500$ Hz ($c_s\approx1.8$ mm/s).

\begin{figure}[t!]
\centering
\begin{subfigure}{0.48\textwidth}
\includegraphics[width=\textwidth]{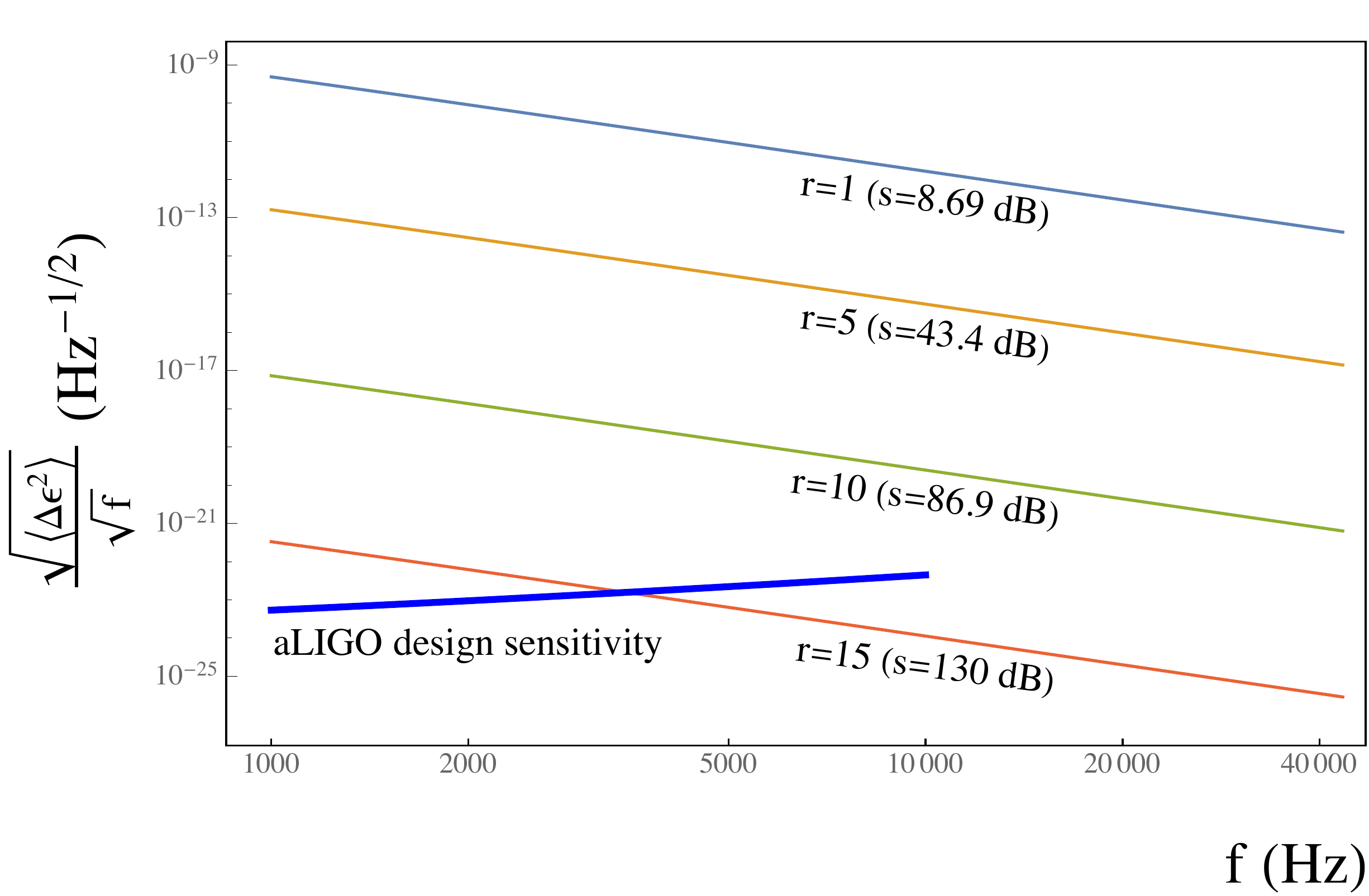}
\caption{}
\label{fig: Future1}
\end{subfigure}\hspace{0.25cm}
\begin{subfigure}{0.48\textwidth}
\includegraphics[width=\textwidth]{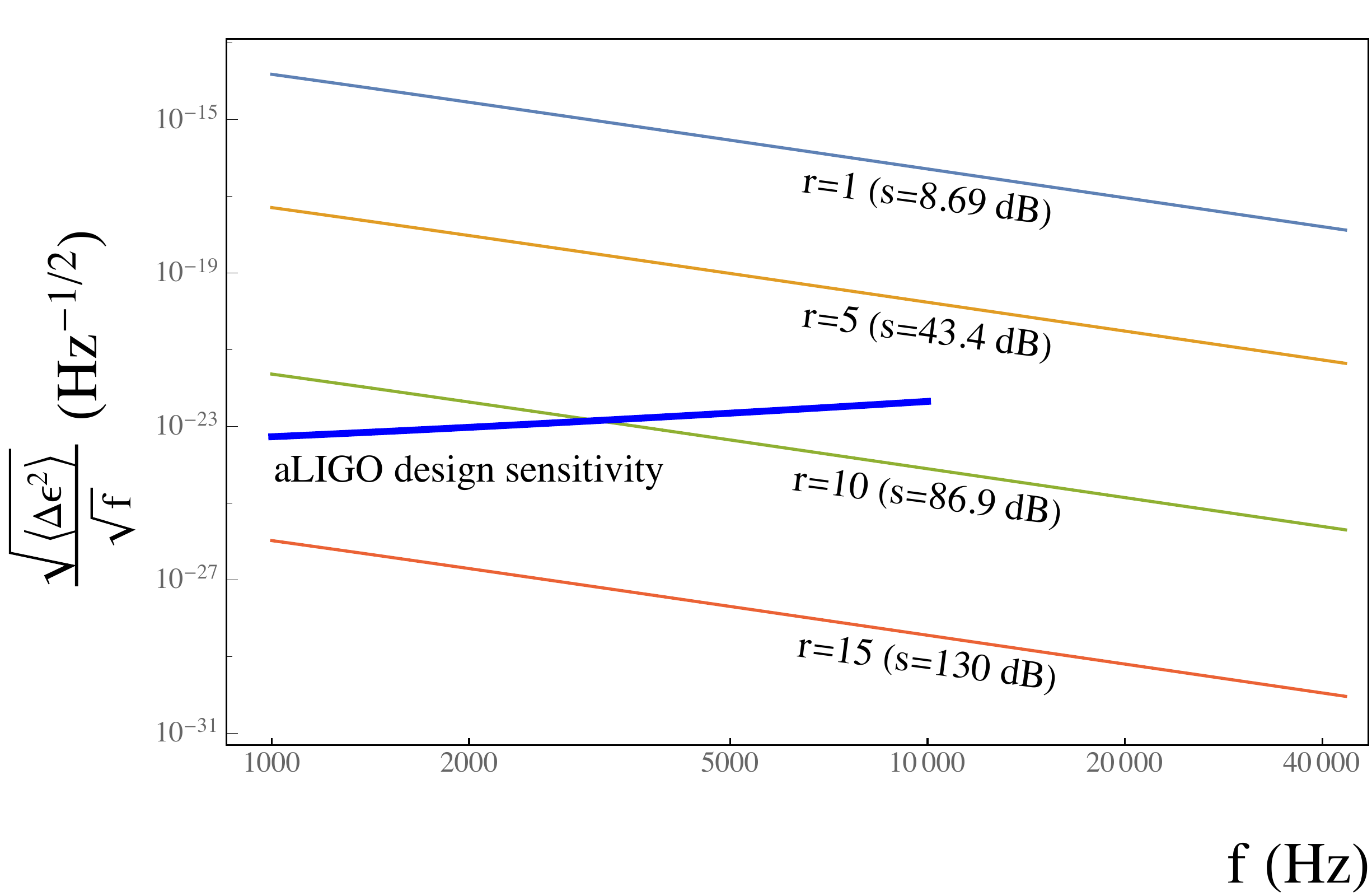}
\caption{}
\label{fig: Future2}
\end{subfigure}
\caption{Sensitivity to kHz gravitational waves for a futuristic BEC with $c_s=1.2\times10^{-2}$ m/s, $t_{obs}=10^6$ s, $\phi=\pi/2$, and $\tau=10^{-3}$ s. In (a), $L=10^{-3}$ m, while $L=1$ m in (b). The thick blue line is the design sensitivity of aLIGO (extrapolated to $10^4$ Hz).}
\label{fig: Future}
\end{figure}

In figure \ref{fig: Future1}, we illustrate how increased squeezing can affect the BEC's sensitivity to gravitational waves. We assume that a BEC in the future can be constructed with similar properties as those in figure \ref{fig: Current 1}, but with $r$ in excess of $0.83$. It is necessary to squeeze phonons above $r=15$ in order to rival LIGO-level sensitivities. In an alternate futuristic scenario shown in figure \ref{fig: Future2}, we suppose that metre-long BECs can be constructed. For this case, we can exceed a LIGO-level sensitivity at $r\approx 10$.  We acknowledge that several difficulties exist in constructing large-scale BECs, such as how to cool to sufficiently low temperatures. This is a major experimental challenge for the future that we will not further consider here. Indeed, attaining values $r \geq 1$ is currently unfeasible, and would require advances in squeezing techniques.

\subsection{Decoherence} 
\label{sec: decoherence}

Let us now look at the maximum value of $\tau$. As derived in \cite{Serafini2005}, in the case that $r_0>\max[\mu_0/\mu_{\infty},\mu_{\infty}/\mu_0]$ the decoherence time of a BEC is
\begin{align}
t_d=\frac{1}{\gamma_B}\ln\left(\frac{\frac{\mu_0}{\mu_{\infty}}+\frac{\mu_{\infty}}{\mu_0}-2\cosh (2r_0)}{\frac{\mu_0}{\mu_{\infty}}-\cosh(2r_0)}\right)\ ,
\label{eq: td}
\end{align}
where $\mu_{0}$ is the initial purity, $\mu_{\infty}$ is the purity as $t\to\infty$, $r_0$ is the initial squeezing (which will decay over time), and $\gamma_B$ quantifies the damping rate. For low temperatures, Beliaev damping is dominant and at zero-temperature is given by \cite{Giorgini1998}
\begin{align}
\gamma_B\approx \frac{3}{640\pi}\frac{\hbar\omega_{\vec{k}}^5}{mnc_s^5}\ ,
\label{eq: gamma}
\end{align}
where $\omega_{\vec{k}}$ is the frequency of the single phonon mode $\vec{k}$, $m$ is the mass of the atoms making up the BEC, and $n$ is the number density.  For simplicity, consider the case in which $r_0$ is large enough such that $\ln\left(\frac{\frac{\mu_0}{\mu_{\infty}}+\frac{\mu_{\infty}}{\mu_0}-2\cosh (2r_0)}{\frac{\mu_0}{\mu_{\infty}}-\cosh(2r_0)}\right)\sim\mathcal{O}(1)$. From Equation (\ref{eq: Delta epsilon continuous}), note that phonons with a frequency $\omega_{\vec{k}}\approx\frac{\Omega}{2}$ are most important for the gravitational wave detection. In this case, sensitivity to gravitational waves is maximized when our measurement duration is in the range $\frac{2\pi}{\Omega}\leq \tau\lesssim\frac{1}{\gamma_B}$.  Taking our BEC to have $c_s=1.2\times10^{-2}$ m/s, $n=7\times10^{20}$ m${}^{-3}$, $m=10^{-25}$ kg,  our sensitivity to gravitational waves is optimal when $\frac{2\pi}{\Omega}\leq\tau\lesssim (3.6\times10^2 {\rm\hspace{0.1em} sec})\left(\frac{f}{2\pi\cdot10^3\rm{\hspace{0.1em}  Hz}}\right)^{-5}$. For a $10$ kHz gravitational wave, we find $t_d\approx3.62$ s, which is greater than the period of the gravitational wave.

We can analyze decoherence effects more rigorously by noting that the squeezing parameter evolves in time as \cite{Serafini2005}
\begin{align}
\cosh[2r(t)]=\mu(t)\left(e^{-\gamma_Bt}\frac{\cosh(2r_0)}{\mu_0}+\frac{1-e^{-\gamma_Bt}}{\mu_{\infty}}\right)\ ,
\label{eq: r}
\end{align}
where
\begin{align}
\mu(t)=\mu_0\left(e^{-2\gamma_Bt}+\frac{\mu_0^2}{\mu_{\infty}^2}\left(1-e^{-\gamma_Bt}\right)^2+2\frac{\mu_0}{\mu_{\infty}}e^{-\gamma_Bt}\left(1-e^{-\gamma_Bt}\right)\cosh(2r_0)\right)^{-1/2}
\end{align}
is the purity. We will now determine the measurement time of $\tau=t$ in order to maximize the sensitivity to gravitational waves. Taking $e^r,e^{r_0}\gg1$ and $\mu_0=\mu_{\infty}=1$, we note that Equation (\ref{eq: r}) behaves as
\begin{align}
e^{2r}\sim\frac{e^{-\gamma_B\tau}e^{2r_0}+1-e^{-\gamma_B\tau}}{\sqrt{e^{-2\gamma_B\tau}+\left(1-e^{-\gamma_B\tau}\right)^2+2e^{-\gamma_B\tau}\left(1-e^{-\gamma_B\tau}\right)e^{2r_0}}}.
\label{eq: r sim}
\end{align}
We see that the squeezing decays from $e^{2r_0}\to e^{r_0}$ on a time-scale of 
\begin{align}
\tau \sim -\frac{1}{\gamma_B}\ln\left[\frac{-\sqrt{e^{2 r_0}-e^{4 r_0}-e^{6 r_0}+e^{8 r_0}}+2 e^{2 r_0}-e^{4 r_0}-1}{4 e^{2
   r_0}-3 e^{4 r_0}-1}\right]\sim\frac{2}{5\gamma_B}
\end{align}
 Consider phonons at a frequency of $\frac{\Omega}{2}$. For $\tau\approx\frac{2}{5\gamma_B}$ and large $e^{r},e^{r_0}$, we can combine Equations (\ref{eq: r sim}) and (\ref{eq: delta epsilon 1 mode}) to find
\begin{align}
\frac{1}{\braket{(\Delta\tilde{\epsilon})^2}}\sim \tau^4 e^{4r_0}e^{-2\gamma_B\tau}\ ,
\end{align}
which is maximized at $\tau=\frac{2}{\gamma_B}$.

Let us now consider more formally how decoherence could affect the sensitivity to gravitational waves. We can incorporate decoherence into (\ref{eq: delta epsilon 1 mode}) with Equation (\ref{eq: r}) by letting $r\to r(t)$ and following the same steps that were used to arrive at Equation (\ref{eq: M}). We also note from \cite{Olivares2012,Howl2018} that the purity divides the covariance matrix, Equation (\ref{eq: sigma 0}), so $\mu(t)$ also multiplies Equation (\ref{eq: delta epsilon 1 mode}). By integrating over all the modes, we can then determine an equation analogous to (\ref{eq: M}), such that $\braket{\Delta \epsilon^2}_{tot}$ now includes effects arising from decoherence.  We note that this is only an approximation; decoherence should, strictly speaking, be introduced prior to solving Equation (\ref{eq: SHO}).

\begin{figure}[t!]
\centering
\includegraphics[width=\textwidth]{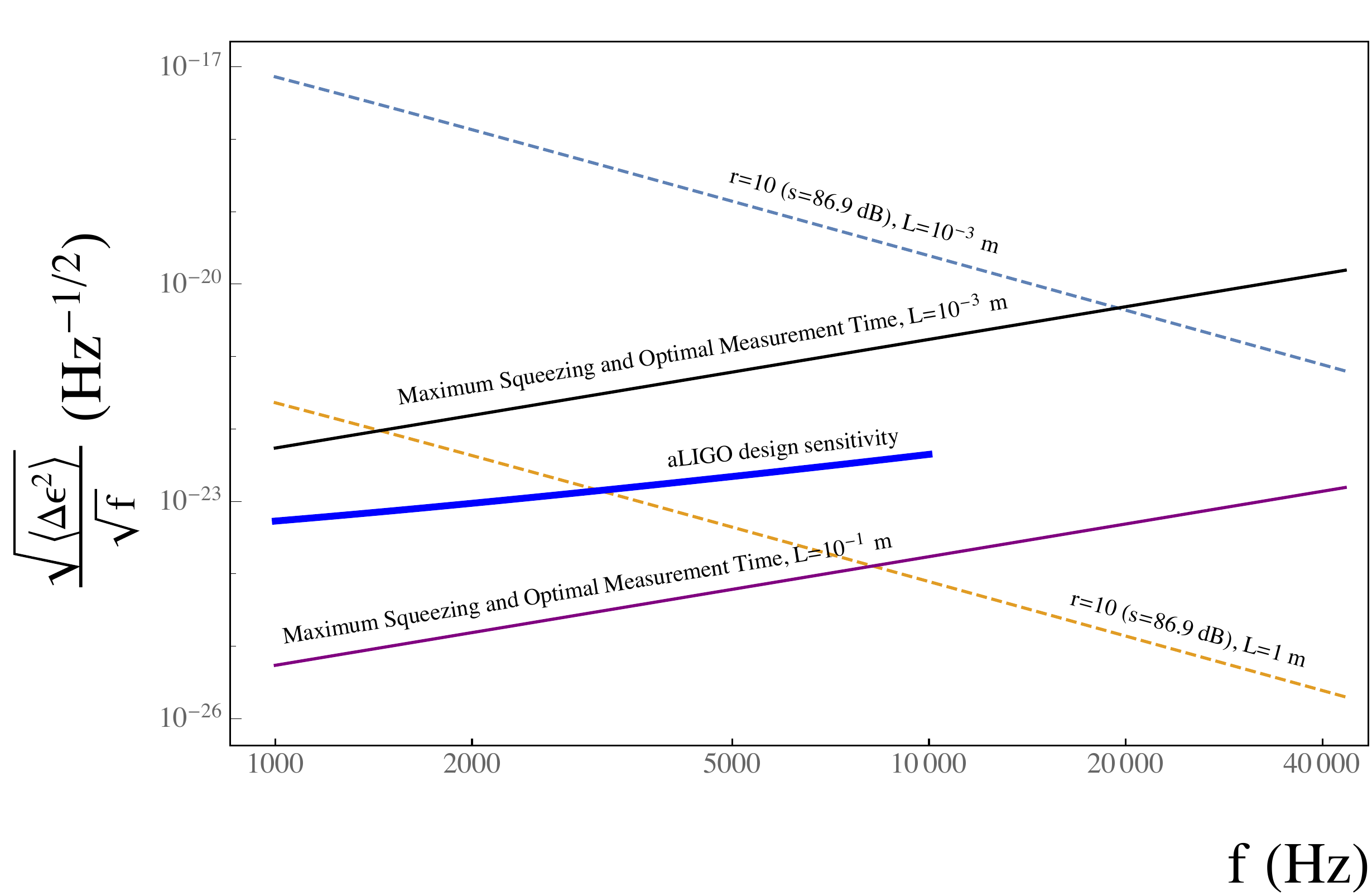}
\caption{ The sensitivity of a BEC to gravitational waves, with $\tau$ optimized and $r_0$ maximized within the validity of the model (solid black and purple lines). Over this frequency range, the measurement duration corresponds to $\tau\sim\frac{1}{\omega}$. We have set $t_{obs}=10^6$ s, $m=10^{-25}$ kg, $n=7\times10^{20}$ m${}^{-3}$, $\mu_0=\mu_{\infty}=1$, $c_s=1.2\times10^{-2}$ m/s, and $\phi=\pi/2$. To facilitate comparison, we have also considered $r=10$ for two different condensate lengths, assuming damping to be negligible (dashed blue and orange lines), where we have $\tau=10^{-3}$ s in both cases. The overall trend of the design sensitivity of aLIGO is indicated by the thick blue line (extrapolated to $10^4$ Hz). 
}
\label{fig: OptimalSqueezing}
\end{figure}

In figure \ref{fig: OptimalSqueezing}, we consider the maximum value of the squeezing parameter from Equation (\ref{eq: approx}) and the optimal measurement duration to maximize the sensitivity to gravitational waves in the kHz range. As we are considering all modes, in this regime $\tau\approx\frac{1}{\omega}$ (verified numerically). We see that, because of decoherence, the sensitivity decreases for higher frequencies.

It will be necessary to constantly regenerate the BEC \cite{Streed2006,Tiecke2009} in order to repeatedly perform measurements over the $t_{obs}=10^6$ s. With such a BEC machine, it will be possible to generate multiple BECs simultaneously. In Equation (\ref{eq: M}), we assumed that the number of measurements was $M\sim t_{obs}/\tau$. For $N$ BECs, the sensitivity to gravitational waves will improve by $1/\sqrt{N}$. To obtain a sensitivity $\frac{\sqrt{\braket{\Delta\epsilon^2}}}{\sqrt{f}}\approx10^{-23}$ Hz${}^{-1/2}$ to a gravitational wave of frequency $f=10^{4}$ Hz using the parameters in figure \ref{fig: Current 1}, we would  therefore require $\mathcal{O}(10^{22})$ BECs, which is impractical. Improved techniques of increasing the squeezing of the phonons and increasing the volume of the condensate will be necessary for a BEC kHz gravitational wave detector to be achievable.

\section{Conclusion}
\label{sec: Conclusion}

We have investigated the feasibility of using a BEC as a gravitational wave detector by modelling the wave as $h_{+}=e^{-t^2/\tau^2}\sin(\Omega t)$, where the exponential prefactor is included to model the measurement duration.
We have derived an analytic expression for the mean-square error in the amplitude of the gravitational wave, which depends on the squeezing of the phonons in the BEC, volume of the BEC, speed of sound, and frequency of the gravitational wave. Turning to a consideration of currently available techniques to improve sensitivity within the linear
dispersion regime,  we find that a BEC constructed using the best possible parameters to maximize the sensitivity will be unable to observe  gravitational waves in the kHz range.    A full analysis would require making use of the non-linear dispersion relation in Appendix \ref{sec: appendix}. 

Though a BEC as a gravitational wave detector is currently not feasible for observing kHz gravitational waves, it could be a promising method for observing waves at this frequency once it is understood how to increase phonon squeezing as well as the volumes of BECs. In the meantime, it will be productive to analyze how properties of the BEC can be optimized to improve sensitivity to gravitational waves by investigating different trap geometries and understanding the effects of vortices and inhomogeneities on the sensitivity. Furthermore, different trapping and optical potentials should be examined as this could have ramifications for the amount of squeezing that can be done and the speed of the phonons. 

While here we focused on BECs at zero temperature, finite temperature effects can further affect the metrology, and decoherence (through Landau damping), and will be studied in future work.  Furthermore, we have considered a homogeneous BEC. Recent work \cite{Schutzhold2018} has emphasized the importance of using inhomogeneous BEC condensates, since they scale with the number of condensate atoms instead of the number of phonons as in the homogeneous case.
It would be of interest to extend our work to the inhomogeneous case to see how better to optimize detection. It is only once these questions are  answered that it will be possible to rival the sensitivities of LIGO (and its successors) for kHz gravitational waves. 
\appendix

\section{Calculation of the BEC's Dispersion Relation}
\label{sec: appendix}

 Let us now determine the general dispersion relation for the BEC. Inserting 
 \begin{align}
 \phi = \sqrt{\kappa^2-m^2 \over 2\lambda} \exp\left[i (\kappa t+\pi) +\sigma \right],
 \end{align}
  (with real $\sigma$ and $\pi$) into Equation \eqref{eq: L}, we find the Lagrangian to be
\begin{align}
\mathcal{L}=\frac{(\kappa^2-m^2)}{2\lambda}e^{2\sigma}\left[\partial_{\mu}\sigma\partial^{\mu}\sigma-(\nabla\pi)^2+(\dot{\pi}+\kappa)^2\right]-\frac{m^2(\kappa^2-m^2)e^{2\sigma}}{2\lambda}-\frac{(\kappa^2-m^2)^2e^{4\sigma}}{4\lambda}\ ,
\end{align}
where
\begin{align}
2\sigma=\ln\left[\frac{(\kappa+\dot{\pi})^2-(\nabla\pi)^2-m^2}{\kappa^2-m^2}\right]\ .
\end{align}
 Writing $\sigma$ and $\pi$ in terms of their inverse Fourier transforms,
\begin{align}
\sigma=\int\hat{\sigma}(\omega,\vec{k})e^{i(k\cdot x-\omega t)}d\omega d^3k\ ,\\
\pi=\int\hat{\pi}(\omega,\vec{k})e^{i(k\cdot x-\omega t)}d\omega d^3k\ ,
\end{align}
we find
\begin{align}
\frac{2\lambda}{\kappa^2-m^2}\mathcal{L}_2=\begin{pmatrix} \pi^* & \sigma^*\end{pmatrix}
\begin{pmatrix}
\omega^2-k^2+2(m^2-\kappa^2) & -2i\kappa\omega \\
2i\kappa\omega & \omega^2-k^2
\end{pmatrix}
\begin{pmatrix} \pi \\ \sigma\end{pmatrix} 
\end{align}
for the quadratic term in the Lagrangian.

Setting the determinant of this matrix equal to zero and solving for $\omega$ gives the dispersion relation,
\begin{align}
\omega^2=k^2-m^2+3 \kappa ^2 \pm \sqrt{m^4+4 k^2 \kappa ^2-6 m^2 \kappa ^2+9 \kappa ^4}\ .
\label{eq: dispersion}
\end{align}
From Equation (\ref{eq: cs}), we can write
\begin{align}
\kappa=\frac{\sqrt{1-c_s^2}}{\sqrt{1-3 c_s^2}}m\ .
\end{align}
Requiring  $k\ll m$ and $c_s\ll1$, Equation (\ref{eq: dispersion}) (for the minus sign, associated with the low-frequency Goldstone mode) can be expanded in $k$ and $c_s$ to find 
\begin{align}
\omega^2 = c_s^2k^2 + {\cal O} \left(\frac{k^4}{m^2} \right) .
\end{align}
Therefore $\omega \approx c_sk$ (indicating that $c_s$ does represent the speed of sound), as long as $\omega \ll \mu \equiv m c^2_s$, i.e. frequency is much smaller than the chemical potential $\mu$.

\section*{Acknowledgements}
We thank Aditya Dhumuntarao, Soham Mukherjee, Carlos Sab\'{i}n, and Joseph Thywissen for useful conversations and correspondence.  We also thank Jorma Louko for catching inconsistencies and helping clarify subtleties in our work. MR was funded by a National Science and Engineering Research Council of Canada (NSERC) graduate scholarship. This research was supported in part by NSERC and Perimeter Institute for Theoretical Physics. Research at Perimeter Institute is supported by the Government of Canada through Industry Canada and by the Province of Ontario through the Ministry of Economic Development \& Innovation.

\bibliography{Bibliography-BEC_Paper_Nov2018}
\bibliographystyle{JHEP}
\end{document}